\documentstyle[12pt]{report}
\begin{document}
\begin{center}
{\large {\bf COMMENT ON ``MEANING OF THE WAVE FUNCTION''}}\\~\\
{\bf J.Samuel and R.Nityananda\\Raman Research Institute\\Bangalore, INDIA
560 080}\\
\end{center}
\baselineskip=24pt
{\sl Abstract}~:~
We draw attention to an elementary flaw in a recently proposed experiment
to measure the wave function of a single quantum system.\\

In a recent paper, Aharonov, Anandan and Vaidman [1] claim to have devised
an experiment to measure the wave function [2] of a {\sl single} quantum
mechanical system. The experiment they propose is a `protective
measurement': while the measurement is being performed, the wave function
is prevented from collapsing by means of suitably chosen external field.
The claim of ref.[1] is at variance with accepted interpretations of
quantum mechanics.  According to these, it is impossible to learn anything
new about a quantum system without disturbing it. Our purpose in this note
is to examine the claim made in ref.[1] more closely.\\

The issues involved are best illustrated by the following simple example.
One can determine the state of polarisation of a classical
laser beam (one with a large number of photons)
by means of simple experiments involving polaroids.
Given a single photon,
conventional wisdom holds that there is no way to measure its polarisation
precisely.  If the photon passes through a polaroid, all one can say with
any certainty is that it was not polarised in the direction orthogonal to
the polaroid.  Ref.[1] challenges this conventional wisdom.\\

Much of reference [1] addresses the question of whether the wave function
has an ``epistemological'' or ``ontological'' meaning.  It is not our intention
to enter into this debate. We restrict our attention to the simplest of
the experiments proposed in [1], which is a modified Stern-Gerlach
experiment. The usual Stern-Gerlach experiment measures the  ${\bf n}$
component of the neutron spin by sending it through an inhomogenous
magnetic field ${\bf B}_1(x)$ along the ${\bf n}$ direction. If the neutron
is polarised parallel to ${\bf n}$, it is deflected in one direction and if
it is antiparallel, it suffers the opposite deflection.  A general spin
state of the neutron is a superposition of these two states. A neutron
polarised in such a state will be deflected one way or another with
probabilities determined by its overlap with the two basis states. A beam
of neutrons polarised in a general direction will split into two beams,
each of which is polarised parallel or antiparallel to ${\bf n}$.  Note
that if the beam was initially polarised along the ${\bf n}$ direction, it
will not split.  The wavefunction does not collapse if it is already in an
eigenstate of the quantity being measured.\\

The modified Stern-Gerlach experiment envisaged in ref.[1] uses an
additional homogeneous, large magnetic field ${\bf B}_0$, which `prevents
the wave function from collapsing'. Ref.[1] claims that with this external
field present, the beam of neutrons does not split and strikes the screen
at one spot, whose location gives us information about the wave function.
To quote ``the beam clearly does not split provided the spin state is
protected by a large homogeneous magnetic field in the {\sl unknown}
direction of the spin''.\\

Ref. [1] gives the impression that the proposed experiment enables one to
measure the quantum state of a single system {\sl without knowing what it
was initially}. To quote ``We do not know what $|\Psi>$ is before the
measurement.'' The purpose of this comment is to point out that one cannot
{\sl learn} anything about the system since the proposed experiment cannot
be performed without prior knowledge of the spin direction. How would an
experimenter create a magnetic field in an {\sl unknown} direction?
Consequently the proposed experiment does not achieve what it claims to
do.\\

We also point out that the additional homogeneous, large magnetic field
${\bf B}_0$ is not necessary in order to carry out a ``protective
measurement''. One can do it by means of a suitably aligned standard
(unmodified) Stern-Gerlach apparatus.  One can `measure' the wave function
of a two-state spin system by physically rotating the Stern-Gerlach
apparatus so that the spin wave function is in one of its eigenstates. But
in order to do this we would need to know the wave function. The situation
in the modified Stern-Gerlach experiment proposed in reference [1] as a
`protected' measurement is no different.  One must know the wave function
{\sl a priori} to measure it. In the language of optics, if a photon is
known to be linearly polarised (let us say it got through a polaroid) it
will certainly go through polaroids aligned with the initial one. This is
the simplest version of a `protective measurement'.\\

If one agrees that the proposed experiment must be a set up without prior
knowledge of the wave function one finds the following well known
situation. Neutrons are passed through a {\sl fixed} magnetic field
${\bf B}({\bf x})$. (The decomposition of ${\bf B}$ into components
${\bf B}_0$ and ${\bf B}_1$ made in ref.[1] is a theoretical construct which
is irrelevant to the behaviour of the  system. The neutron responds to the
total magnetic field in which it is placed). For any fixed choice of the
magnetic field ${\bf B}(x)$, there exist spin states $|a>$ and $|b>$ (the
protected ones, in the language of reference [1]) of the neutron which
lead to the neutron being incident with probability one at points A and B
on the screen respectively.  Any other spin state $|\psi>$ is a linear
combination of $|a>$ and $|b>$ and from the superposition principle, will
strike the screen at A {\sl or} B with probabilities given by
$|<\psi|a>|^{2}$ and $|<\psi|b>|^{2}$. If A and B are distinct, a neutron
beam in a general spin state {\sl will} split. On the other hand, if A and
B are the same, all states will strike the screen at the same spot.  It is
easy to construct magnetic field configurations for which A and B
coincide.  One example is a homogeneous magnetic field.  A more general
example is any field [6] which has vanishing ${\nabla}|{\bf B}|$.  For
these magnetic fields, all states strike the screen at the same spot,
whose location tells us nothing whatever about the initial state. This
experiment would then be a protective {\sl non} measurement.\\

In summary, any experiment that increases our knowledge of the state of a
quantum system necessarily disturbs it. If one does not disturb the system
one learns nothing more about it. These are the basic features of quantum
measurement [7] which were challenged in reference [1]. The usual
interpretation of the wave function is thus protected.

\newpage
\noindent
{\bf References}\\
\begin{enumerate}
\item Y. Aharonov, J.Anandan and L.Vaidman, {\sl Phys. Rev. A}, {\bf 47},
4616 (1993). Related references are [3] and [5].

\item More precisely, the wave function up to an overall phase. In the
following we ignore this relatively fine distinction between a quantum
mechanical state and a ray.

\item David Freedman ``Theorist to the quantum mechanical wave: Get Real''
{\sl Science}, {\bf 259}, 1542 (1993).

\item Reference [1] has also been criticised for the same basic reason in
[5] . However, we disagree on salient points with [5]. To minimise
confusion, we restrict our remarks here to ref.[1].
\item Carlo Rovelli, High Energy Physics Bulletin board (hep-Th/9304126).

\item In the example discussed in [1] ${\bf B} = {\bf B}_0 +
{\bf B}_1(x)$, where $|{\bf B}_0|$ is much larger than
$|{\bf B}_1|$, this situation obtains if ${\bf B}_0$ is orthogonal
to ${\bf B}_1$.

\item P.A.M.Dirac Principles of Quantum Mechanics, (OUP, London 1958)
\end{enumerate}
\end{document}